\documentstyle[12pt]{article}

\begin{document}
\renewcommand{\thefootnote}{\fnsymbol{footnote}}
\begin{flushright}
University of Adelaide preprint: \\
ADP-95-60/T205 \\
(to appear: Mod. Phys. Lett. {\bf 1996})
\end{flushright}
\vspace{1.0cm}
\begin{center}
{\huge \bf A New Hypothesis on the Origin of the 
Three Generations \\}
\vspace{3ex}
\vspace{3ex}
{\large \bf S. D. Bass
\footnote{sbass@cernvm.cern.ch}$^{[1,2,4]}$ 
and A. W. Thomas
\footnote{athomas@physics.adelaide.edu.au}$^{[3,4]}$ \\}
\vspace{3ex}
{\it $^1$ Cavendish Laboratory, University of Cambridge, Cambridge
CB3 0HE, U.K. \\}
\vspace{3ex}
{\it $^2$ Institut f\"ur Kernphysik, KFA-J\"ulich, D-52425 J\"ulich, Germany 
\\}
\vspace{3ex}
{\it $^3$ Department of Physics and Mathematical Physics, University of
Adelaide, \\
Adelaide 5005, Australia \\}
\vspace{3ex}
{\it $^4$ Institute for Theoretical Physics, University of Adelaide, 
Adelaide 5005, Australia \\}

\vspace{3ex}

\vspace{3ex}
{\large \bf ABSTRACT \\}
\end{center}
\vspace{3ex}
{
We suggest that the Standard Model may undergo a supercritical 
transition near the Landau scale,
where the U(1) gauge boson couples to the left and right handed
states
of any given fermion with different charges.
This scenario naturally gives rise to three generations of 
fermion, corresponding to the three critical scales
for the right-right, right-left and left-left fermion
interactions going supercritical, as well as
CP violation in the quark sector. }

\newpage

\section{Introduction}

The Standard Model has proven very successful in
every area of particle physics, including recent 
high-energy collider experiments -- e.g. see Ref.\cite{1}.
However, it has three features which are not well understood:
the origin of mass, the three fermion generations and the phenomenon of
CP violation.
The question of mass is usually framed in terms of (fundamental) Higgs
fields \cite{Higgs} and why
the corresponding Yukawa couplings take particular values 
-- see however Ref.\cite{2}. Instead, we believe that one should
ask whether a formulation of the Standard Model with massless fermions 
makes sense. For example, it is well known that QED
with massless electrons
is not well defined at the quantum level \cite{Muta,8}.

In this paper we consider the pure Standard Model with gauge symmetry
$SU(3) \otimes SU(2)_L \otimes U(1)$ and no additional interaction;
that is,
assuming no additional unification.
We examine the physical theory that corresponds to the bare Standard
Model Lagrangian
with no elementary Higgs and just one generation of fermions and gauge
bosons which all
have zero bare mass.
At asymptotic scales, where the U(1) coupling is significantly
greater
than
the asymptotically free SU(3) and SU(2)$_L$ couplings,
the left and right handed states of any given charged fermion couple to
the U(1) gauge boson
with different charges.
At the Landau scale there will be three separate
phase transitions corresponding to each of the right-right, right-left
and left-left interactions becoming supercritical. These transitions
correspond to three generations of fermions. As one passes through each
transition from a higher scale (shorter distance) the corresponding
scalar condensate ``melts'', releasing a dynamical fermion into the
Dirac phase studied in the laboratory. In this picture the three
generations emerge as quasi-particle states built on a ``fundamental
fermion'' interacting self-consistently with the condensates.

Clearly this proposal differs in a fundamental manner from the
conventional approaches to the Standard Model. While the conceptual
framework is extremely simple and 
elegant, the techniques for dealing with
non-perturbative physics at the Landau scale are not well developed.
In particular, at the present stage we are not able to present a
rigorous, quantitative derivation of all of the features of the Standard
Model. Nevertheless, we believe that the potential for understanding so
many phenomena, including mass, CP-violation and the generations, is so
compelling that the ideas should be presented at this stage.

The structure of the paper is the following.
In order to introduce the ideas we first review the change in the vacuum
of pure QED near a point-like nucleus with charge greater than $137$,
a problem that has received enormous effort \cite{Grein1,Grein2,Orsay}. We then
consider the analogous case of QED at the Landau scale, from which we
conclude that in pure QED the electron would self-consistently generate
its own mass. Having thus introduced the basic notions we turn to the
full Standard Model, considering in turn the generations of charged
fermions and the origin
of CP violation, the neutrinos and the mass of the vector bosons.

\section {The Supercritical Phase of Non-Asymptotically Free Gauge Theory }

Several decades of work have revealed that
non-asymptotically free (NAF) U(1) gauge theories
with zero bare fermion mass
are capable of generating their own renormalised mass.
The prime example of such a theory is, of course, QED with zero bare mass,
where both analytic \cite{mas,fk,Miran} and numerical (lattice) 
calculations \cite{Latt} have shown
that one finds a finite renormalised mass and a non-trivial, 
ultraviolet (UV), stable fixed point.
Without this the theory is trivial; that is,
the charge is completely screened by the
interactions so that the theory is equivalent to a free field theory.

The possibility that QED may be trivial was originally 
suggested by Landau and co-workers \cite{Land1,Land2}
and Fradkin \cite{Frad}.
Consider the runnning coupling in perturbative (weakly coupled) QED.
The one loop vacuum polarisation implies
\begin{equation}
\alpha (m^2) = { \alpha (\lambda^2) \over 1 + {\alpha (\lambda^2) \over 3 \pi}
\ln { \lambda^2 \over m^2 }	}.
\end{equation}
If we take $\lambda^2 \rightarrow \infty$ (the continuum limit of 
QED with a finite cut-off)
then $\alpha (m^2)$ vanishes for all $m^2$.
The same result applies in the limit of zero mass gap
(i.e. $m^2 \rightarrow 0$),
namely the coupling $\alpha (m^2)$ vanishes again. 
Recent work by Koci\'c et al. \cite{Kocic} has shown that
this ``zero charge problem'' persists
when the magnetic interaction of the electron is also included, despite 
the fact that it tends to screen the vacuum polarisation.

Another indication of the problem of massless QED is the fact that 
one cannot renormalise it (perturbatively) on mass shell ( which is
a necessary condition for the electron to be a physical particle).
The only alternative, which was once again suggested by 
Landau et al. \cite{Land1} (see also Dirac \cite{Dirac}),
is that non-perturbative effects near the Landau scale mean that QED has
a non-trivial, UV, stable fixed point. 
A number of groups \cite{mas,fk,Miran,Miran2} have shown 
that QED, in quenched, ladder approximation, has
a non-trivial, UV, stable
fixed point at $\alpha_c = {\pi \over 3}$,
which separates
the weakly and strongly interacting phases.
The theory is trivial for bare coupling $\alpha < \alpha_c$, whereas for
$\alpha > \alpha_c$
the chiral symmetry of the massless bare theory is spontaneously broken by
the interactions leading to the formation of tightly bound states -- 
much like the $Z > 137$ point nucleus problem in QED.
Kogut et al. \cite{Latt} have found that this UV, stable fixed point 
survives in unquenched lattice QED.
Estimates of the value of $\alpha_c$ (the critical bare coupling)
from Schwinger-Dyson and lattice calculations range between
0.8 and 2 [18-21].

Given that a NAF U(1) gauge theory has a non-trivial, UV, stable
fixed point, $\alpha_c$,
it follows that the theory has a two phase structure.
We let $\lambda_c$ denote the scale at which $\alpha$ reaches the fixed point
$\alpha_c$
and call the phases at scales above and below the critical scale $\lambda_c$
the Landau and Dirac phases respectively.
Perturbative QED (and the Standard Model) is formulated entirely in the Dirac
phase of theory ($ \mu < \lambda_c $).
The theory seems to behave as a gauged Nambu-Jona-Lasinio model
\cite{NJL} in the Landau phase \cite{Latt,Bard,Miran3}.

The connection with supercritical phenomena (in particular, the large-$Z$,
point-nucleus problem)
suggests a simple physical interpretation of this theory.
Since massless, perturbative QED is not a consistent theory because of
the ``zero charge problem'',
we consider perturbative QED with a finite renormalised mass and sketch 
how this mass could be recovered self-consistently in a complete formulation
of QED.

The coupling $\alpha$ increases until we reach the critical
scale $\lambda_c$ where the interaction of the fermions 
with the gauge field becomes supercritical.
To understand what happens at this transition 
it is helpful to consider the analogous problem of a
static, large-$Z$, point nucleus in QED \cite{Grein1,Grein2,Orsay}.
There the $1s$ bound state level for the electron falls into the 
negative energy continuum at $Z=137$.
If we attempt to increase $Z$ beyond 137 the point nucleus becomes a
resonance: an electron moves from the Dirac
vacuum to screen the supercritical charge 
which then decays to $Z-1$ with the emission
of a positron.

If the electron itself were to acquire a supercritical charge at
very large scales, {\cal O}$(\lambda_c)$,
it would not be able to decay into a positive energy bound state
together with another electron because of energy momentum conservation.
In this case, the Dirac vacuum itself would decay to a new
supercritical vacuum state. Since the vacuum is a scalar,
this transition necessarily involves the formation 
of a scalar condensate 
which spontaneously breaks the (near perfect) 
chiral symmetry of perturbative QED at large momenta.
The Dirac vacuum
is a highly excited state at scales $\mu \geq \lambda_c$
and one must re-quantise the fields
with respect to the new ground state vacuum in the Landau phase of the theory.
The Dirac electron of perturbative QED would freeze
out of the theory as a dynamical
degree of freedom and the running coupling
would freeze at $\alpha (\lambda_c)$.
Perturbative QED, which requires a finite electron mass, is formulated 
entirely in the Dirac phase of the theory.
The normal ordering mismatch between the zero point energies of the scalar
vacua in the Dirac and Landau phases of QED 
means that the electron in the Dirac phase always feels a uniform, local,
scalar potential. This potential must be included in
the Hamiltonian for perturbative QED. The minimal
gauge invariant, local, scalar operator that we can construct 
is the scalar mass term $m_e [ {\overline e} e ]$.
A self-consistent treatment of QED appears to generate its own mass.

\section {Generations in the Standard Model}

We now discuss how the considerations of the previous section carry over
to the Standard Model.
The Standard Model
differs from QED at very large momentum in that the U(1) gauge boson coupling
to a fermion depends on its chirality.
The right-right, right-left and left-left fermion interactions have
different strengths for the Dirac leptons ($e, \mu$ and $\tau$) 
and the quarks. As we now explain,
this important difference means that a non-perturbative solution
of the Standard Model requires three generations of fermions.
The fermion gauge boson interaction in the electroweak sector is described
by the Standard Model Lagrangian
with symmetry $SU(3) \otimes SU(2)_L \otimes U(1)$:
\begin{equation}
{\overline \Psi}_L \Biggl( \hat{\partial} - i g_1 \hat{B} - i g_2
\hat{\underline{W}}.{1 \over 2}
{\underline \tau} \Biggr) \Psi_L
+ {\overline \Psi}_R \Biggl( \hat{\partial} - i g_1 \hat{B} \Biggr) \Psi_R.
\end{equation}
Here 
$\Psi_L$ and $\Psi_R$ include the left and right handed fermions according to
the Standard Model. We
use $\alpha_1 = {g_1^2 \over 4 \pi}$, $\alpha_2 = {g_2^2 \over 4 \pi}$ and
$\alpha_s$ to denote the U(1), SU(2) and colour SU(3) couplings respectively.

Consider the Standard Model evolved to some very large scale, much greater 
than the ``unification scales" where
the U(1) coupling $\alpha_1 = \alpha_2$ and $\alpha_1 = \alpha_s$.
As we approach the Landau scale
the $Z^0$ evolves to become the U(1)
gauge boson as
$\sin^2 \theta_W \rightarrow 1$.
Since the SU(2) and SU(3) sectors of the Standard Model 
are asymptotically free \cite{GG}
the $W^{\pm}$, the photon and the gluon have effectively disappeared
at these scales.
The $Z^0$ mass increases logarithmically with increasing $\mu^2$ and
can be treated as negligible
at the Landau scale so that the theory behaves as a U(1) gauge field coupling
to left and right handed
fermions with different charges.
The $Z^0$ coupling to the fermions is \\
\begin{equation}
-i g_1 \gamma_{\mu}
\Biggl( c_L {1 - \gamma_5 \over 2} + c_R {1 + \gamma_5 \over 2} \Biggr)
\end{equation}
where
the left and right handed charges $c_L g_1$ and $c_R g_1$ are given in Table 1.
(Here $l$ denotes the charged leptons and $\nu_l$ the corresponding
neutrinos. We use 
$q^{*}$ and $q_{*}$ to denote the upper and lower 
components of the electroweak quark doublet.)

\begin{table}
\begin{center}
\caption{The fermion couplings to the $Z^0$.}
\begin{tabular} {ccc}
\\
\hline\hline
\\
	& $c_L$ & $c_R$ \\
\\
\hline
\\
$l$	& $ - {1 \over 2} + \sin^2 \theta_W $ & $ \sin^2 \theta_W$ \\
\\
$\nu_l$	& $ + {1 \over 2} $ & 0 \\
\\
$q^{*}$	& $ + {1 \over 2} - {2 \over 3} \sin^2 \theta_W $ & 
$ - {2 \over 3} \sin^2 \theta_W$ \\
\\
$q_{*}$   & $ - {1 \over 2} + {1 \over 3} \sin^2 \theta_W $ &
$ + {1 \over 3} \sin^2 \theta_W$ \\
\\
\hline\hline
\end{tabular}
\end{center}
\end{table}

The idea that we wish to develop is the following.
Consider a ``fundamental fermion",
which
is defined in the pure Landau phase of the Standard Model.
Since the left and right handed charges have 
different values, it follows that the
left-left, left-right and right-right
fermion interactions will, in general, become sub-critical at different scales
as we evolve the theory
through the supercritical transitions to lower $\mu^ 2$.
The first interaction to become
sub-critical as we decrease $\mu^2$ is the 
left-left interaction, followed by the 
left-right
and then the right-right interactions.
Each transition is associated with the melting 
of a scalar condensate which releases
a dynamical fermion
into the Dirac phase of the Standard Model.
These Dirac fermions interact self-consistently with
the condensates in the Landau phase of the theory.
In this picture
the three fermion generations emerge as three quasi-particle states 
in the Dirac phase 
which correspond to the ``fundamental fermion" 
in the Landau phase and which couple
to the gauge field with identical charge.

Let us now outline how this structure should be manifest from the
opposite direction, as we
evolve the Standard Model upwards
from the ``low scale" of the laboratory towards the Landau scale.
For simplicity, we first consider the charged leptons.
In the absence of any other physics the Standard Model 
should undergo a rich series of phase transitions near
the Landau scale
as each of the right-right, right-left
and left-left fermion interactions become supercritical with increasing
$\mu^2$. These transitions 
can be classified into one of two types: ``static" transitions and ``vacuum"
transitions. ``Static" transitions
involve the decay of the left or right handed component of a ``heavy" fermion
$\Psi_h$ into a ``light" fermion $\Psi_l$ together with 
the formation of a $(\Psi_h {\overline \Psi}_l)$ bound state
(like the decay of a large-$Z$ point nucleus).
The supercritical component of $\Psi_h$
becomes a resonance
between the critical scales for the static transition and the vacuum transition
at which the $\Psi_h$ freezes into the Landau phase.
``Vacuum" transitions involve the decay of the fermionic vacuum from the Dirac
into the Landau phase and the formation of a scalar condensate.
Static transitions
do not affect the symmetry or generation structure (which is given by the
vacuum transitions).
They do affect the scale at which the vacuum transition involving $\Psi_h$
takes place. The charge of the
``resonance fermion" increases more slowly with
increasing $\mu^2$ than we would
predict using perturbative arguments alone so that vacuum transitions which
involve the resonance fermion are pushed to a higher scale.

At very large scales where
$\sin^2 \theta_W \rightarrow 1$, the charges of the left 
and right handed charged 
leptons become 
$c_L \rightarrow {1 \over 2}$ and $c_R \rightarrow 1$ respectively.
The interaction between two right handed
fermion fields
(eg. $e^-_R$, $e^+_L$) is the first to go supercritical.
One finds the static decays
of the right handed muon $\mu^-_R$ and tau $\tau^-_R$, viz.
$\tau^-_R \rightarrow (\tau^-_R e^+_L) \ e^-_R$,
and also the vacuum transition involving
the right handed electron at a
critical scale $\lambda_c^{RR}$.
Since the vacuum is a scalar, this vacuum transition must be associated with 
the formation of a scalar condensate.
It is important to consider
what has happened to the left handed electron at this point.
The Dirac vacuum for the left-handed electrons collapses at $\lambda_c^{RR}$
because of the axial anomaly \cite{ABJ},
whereby the chirality of a charged
lepton in the Dirac phase is not conserved in the presence of a 
background gauge field.
The anomaly has a simple interpretation in a two phase NAF gauge theory 
\cite{Bud}. Consider
the gauge-invariant axial-vector current in perturbative QED with an explicit 
UV cut-off, which we shall take to be equal to $\lambda_c^{RR}$.
The anomaly appears as a flux of chirality (or spin) over
the cut-off -- and into the Landau
phase of the theory.
If one turns off the anomaly, the Dirac vacuum for the left handed electrons is
highly excited with respect
to the Landau vacuum for the right handed electrons at 
$\mu \geq \lambda_c^{RR}$.Via the axial anomaly, the left-handed electrons  
condense with the right-handed electrons to form the Landau vacuum 
which is created at $\lambda_c^{RR}$
and the electron completely freezes out of the theory. 

At scales $\mu \geq \lambda_c^{RR}$ the remaining charged lepton degrees of 
freedom are the $\mu$ and the $\tau$.
Here the right handed muons and taus
are supercritical resonances while the left handed muons and taus are
still perturbative fermions.
The left handed charge evolves significantly faster than the right handed
charge with increasing
$\mu^2$ and the left handed fermions drive the dynamics.
The muon freezes into the Landau phase at the critical scale
$\lambda_c^{LR}$ for the left-right vacuum transition 
and the tau freezes out 
at the left-left vacuum transition.
The latter is catalysed by the axial anomaly in the same way as the right-right
vacuum transition.
The three self-supercritical transitions (right-right, left-right and 
left-left) yield 
three condensates in the Landau phase of the Standard Model.

The same arguments hold in the quark sector but there is one 
important new point to note.
The upper and lower components of the electroweak quark doublet 
$q^{*}$ and $q_{*}$ become 
self-supercritical at different scales because of the different coupling of
the U(1) gauge boson
to each of the $q^{*}$ and $q_{*}$ quarks. 
This means
that the eigenstates of the $W^{\pm}$-quark interaction in the
Standard Model (which
define the components of the quark doublet)
and the quark mass eigenstates are not identical.
The three generations of quarks mix according to a unitary 
(Kobayashi Maskawa) matrix which, in general, 
gives CP violation in the quark sector.
To see that we have a CP violating interaction at large scales, consider
the vector, vector, axial-vector triangle diagram.
This is anomaly free in the pure Dirac phase of the Standard Model when 
we sum over $l$, $\nu_l$, $q^{*}$ and $q_{*}$
propagating in the triangle loop.
At intermediate momentum scales, where one component 
of the quark doublet has frozen
into the Landau phase
and the other component remains in the Dirac phase, there is a nett 
three-gauge-boson contact interaction
in the Dirac phase of theory which carries the CP-odd
quantum numbers of the axial anomaly.
This corresponds to regularising the UV 
behaviour of the triangle amplitude
with a slightly different
cut-off for each component of the electroweak doublet in perturbation theory. 
Since this cut-off is so much greater than any mass 
scales that are currently amenable to experiment
this contact interaction does not harm either anomaly cancellation or
the renormalisability of the Standard Model.

As the right-handed (Dirac) neutrino is non-interacting in the Standard Model
we cannot see how to form a scalar, neutrino condensate at the
supercritical transitions. Of course, each type of neutrino will sense
the corresponding charged lepton transition 
(through the coupling $\nu_l \rightarrow W l \rightarrow \nu_l$).
While it may be that this coupling gives rise to a
mass, $\nu_l$, of the order of $G_F m_l$ in the Dirac phase, it seems most
likely that the neutrinos are massless.
(The Standard Model with massive gauge bosons and massless neutrinos can be
renormalised on mass shell \cite{aoki}.)
If this is the case it is trivial that there should be no
Kobayashi-Maskawa matrix in the lepton sector:
one can simultaneously diagonalise 
the eigenstates of mass and the $W^{\pm}$-lepton interaction.

The resonance structure offers a possible reason why the top quark is so much
heavier than the bottom quark.
The relative separation of the left-right and left-left
transitions
is greater for the $q^*$ (top quark) than the $q_*$ (bottom quark).
This means that the
top quark has further to evolve than the bottom quark to get from
$\lambda_c^{RL}$
to $\lambda_c^{LL}$
with a slowly increasing left handed charge;
the top quark freezes out at a much higher scale than the bottom quark
and has a much higher mass.
Similarly, the charm quark has a lot further to go than the strange quark
between the right-right and left-right transitions.

The dynamical chiral symmetry breaking which gives us the fermion masses
also gives mass to the gauge bosons.
The gauge fixing in the ``fundamental'' bare Lagrangian (with zero mass)
does not involve
the $0^{-+}$ Goldstone bosons,
which are generated
with the dynamical chiral symmetry breaking when we turn on the vacuum 
polarisation.
The propagators for the gauge bosons are transverse in covariant (eg. Landau)
gauge:
\begin{equation}
\Pi_{\mu \nu} = f^2 \Biggl( g_{\mu \nu} - {p_{\mu} p_{\nu} \over p^2} \Biggr)
\end{equation}
When we evaluate the gauge boson self-energies using the Schwinger-Dyson
equations
(eg. 
in leading logarithm approximation \cite{top})
we find a non-transverse mass term in $\Pi_{\mu \nu}$ which is proportional to
$g_{\mu \nu}$.
The transversity of $\Pi_{\mu \nu}$ is restored by the mixing of 
the gauge bosons
with the $0^{-+}$ Goldstone bosons.
The fermion, gauge-current vertex becomes:
\begin{equation}
\Biggl( \gamma_{\mu} {1 \over 2} (1 - \gamma_5) {\tau^a \over 2} \Biggr)_{{\rm 
bare}}
\rightarrow
\Biggl( \gamma_{\mu} {1 \over 2} (1 - \gamma_5) {\tau^a \over 2} - 
f_{ab} g_b {p_{\mu} \over p^2}	\Biggr)_{({\rm Standard \ Model})}
\end{equation}
where
$f_{ab}$ is the current-Goldstone transition amplitude and $g_b$ 
denotes the Goldstone-fermion coupling.
The Higgs mass and the Goldstone parameters $f_{ab}$ and $g_b$ 
are determined 
by the mass of the top quark $m_t$ and the running QCD coupling
$\alpha_s(m_t^2)$.
As Gribov has emphasised \cite{top},
the Schwinger-Dyson equations for the Higgs and Goldstone self-energies
involve all the fermions on an equal footing.
The top quark becomes important only because of its large mass; it has no
special interaction.

\section {Conclusions}

We have argued, on quite general grounds, that in the absence of
elementary Higgs (or other, additional, physics) the Standard Model 
may generate its own mass and three generations of fermions as a result of
super-critical phenomena at the Landau scale.
We have not considered gravity and one may worry that,
at least in a perturbative treatment,
the Landau scale is larger than the Planck mass.
However, we believe that the scenario presented in this paper is compelling
and certainly merits further investigation.
One could speculate that in a non-perturbative treatment the physics of the
Planck scale and the Landau scale
may in fact be coupled.

It is clearly important to explore the physics of the Landau scale
in the laboratory. This is difficult in the U(1) sector
because of the large momentum scales involved. On the other hand,
in QCD the Landau scale is in the infra-red and it might be that one can learn
a little about
the mechanism proposed here through the study of phenomena such as quark
confinement and hadronisation [30-33].

\vspace{2.0cm}

{\large \bf Acknowledgements \\ }

We would like to thank T. Goldman, V. N. Gribov, 
P. A. M. Guichon, C. A. Hurst, R. G. Roberts, D. Sch\"utte, 
R. Volkas and A. G. Williams for helpful discussions.
We would also like to thank J. Speth for his hospitality at the KFA
J\"ulich where this work began.
This work was supported in part by the Australian Research 
Council and the Alexander von Humboldt Foundation.


\end{document}